Accuracy of a Large Language Model in Distinguishing Anti- And Pro-vaccination Messages on

Social Media: The Case of Human Papillomavirus Vaccination


Soojong Kim[1], Kwanho Kim[2], Claire Wonjeong Jo[1]

[1] Department of Communication, University of California Davis, United States

[2] Department of Communication, Cornell University, United States



**Author Note**

Soojong Kim  https://orcid.org/0000-0002-1334-5310

Correspondence concerning this article should be addressed to Soojong Kim, 1 Shields

Ave, Kerr Hall #361, Davis, CA 95616, United States. Email: sjokim@ucdavis.edu





## Abstract

**Objective**. Vaccination has engendered a spectrum of public opinions, with social media acting as a crucial platform for health-related discussions. The emergence of artificial intelligence technologies, such as large language models (LLMs), offers a novel opportunity to efficiently investigate public discourses. This research assesses the accuracy of ChatGPT, a widely used and freely available service built upon an LLM, for sentiment analysis to discern different stances toward Human Papillomavirus (HPV) vaccination.

**Methods**. Messages related to HPV vaccination were collected from social media supporting different message formats: Facebook (long format) and Twitter (short format). A selection of 1,000 human-evaluated messages was input into the LLM, which generated multiple response instances containing its classification results. Accuracy was measured for each message as the level of concurrence between human and machine decisions, ranging between 0 and 1.

**Results**. Average accuracy was notably high when 20 response instances were used to determine the machine decision of each message: .882 ($SE$ = .021) and .750 ($SE$ = .029) for anti- and pro-vaccination long-form; .773 ($SE$ = .027) and .723 ($SE$ = .029) for anti- and pro-vaccination short-form, respectively. Using only three or even one instance did not lead to a severe decrease in accuracy. However, for long-form messages, the language model exhibited significantly lower accuracy in categorizing pro-vaccination messages than anti-vaccination ones.

**Conclusions**. ChatGPT shows potential in analyzing public opinions on HPV vaccination using social media content. However, understanding the characteristics and limitations of a language model within specific public health contexts remains imperative.








Accuracy of a Large Language Model in Distinguishing Anti- And Pro-vaccination Messages on

Social Media: The Case of HPV Vaccination

Vaccination continues to be a subject of intense public discussion, with a broad spectrum

of viewpoints and beliefs, ranging from advocates praising its benefits to a skeptical faction [1–3].

Given that these diverse perspectives are tied to individuals' health behaviors, understanding

public perceptions of vaccination is of great importance for social scientists and public health

professionals [1,4,5].

As digital platforms, particularly social media, have emerged as pivotal venues for

discussions on health-related issues, researchers have turned to analyzing messages on these

platforms to gain insights into public perceptions [6–9]. At the core of various quantitative and

computational approaches exploring the immense volume of online messages generated on these

platforms lies the process of human evaluation. Often, multiple researchers or experts assess a

subset chosen from a large dataset of online messages, and the insights drawn from the subset are

then extrapolated to the entire dataset or to the broader population through statistical assumptions

or machine-learning techniques [10–12]. However, the human evaluation process is inherently time-

consuming and labor-intensive, demanding extensive collaboration among multiple individuals.

Recently, the advent of large language models (LLMs) has opened up new possibilities to

reduce the burdens associated with human evaluation. LLMs, such as OpenAI's GPT (Generative

Pre-trained Transformer) and Google's LaMDA (Language Model for Dialogue Applications),

are artificial intelligence models trained on a large volume of text data to generate human-like

text based on the user input they receive [13,14]. LLMs have demonstrated considerable capacity for

human-level decision-making and logical processing [15,16]. Furthermore, the increasing



accessibility and user-friendliness of these powerful LLMs are amplifying their impacts in various academic disciplines [17,18].

Therefore, the present research explores the feasibility of utilizing an LLM in investigating public perceptions based on digital platform data. Our primary focus is on ChatGPT, with a particular emphasis on its freely available and most unrestrained iteration powered by GPT 3.5 [19]. ChatGPT powered by GPT 3.5 also distinguishes itself as the most widely used chatbot service with over 100 million monthly users worldwide [20], while operating on one of the most high-performing LLMs available [21]. These characteristics underscore its potential as a feasible and effective tool accessible to a broad spectrum of researchers, including those without substantial financial resources and technical knowledge.

We evaluated the accuracy of ChatGPT operating on GPT 3.5 in classifying the stances toward vaccination expressed in social media messages, by utilizing multiple datasets and comparing human and machine evaluations of the same data. Through this investigation, we aim to contribute to identifying methodological advances for researchers in the fields of public health and social sciences, ultimately enhancing our understanding of public perceptions of health-related issues in the digital era.

Among various issues that stimulate intense public discussion on vaccination, we focused on Human Papillomavirus (HPV) vaccination. Despite its pivotal role as a preventive measure against a spectrum of cancers [22], HPV vaccination encounters significant resistance and skepticism [23,24]. Understanding public perceptions about HPV vaccination and investigating different beliefs that influence its acceptance or resistance is thus a public health priority.

**Method**

**Data Collection**



We retrieved messages related to HPV vaccination from two major social media platforms supporting different message formats: Facebook (long format) and Twitter (short format). Specifically, 141,479 messages were collected from Facebook, and employing the same search criteria used for Facebook, 676,193 messages were obtained from Twitter. This research was exempted by the Institutional Review Board of [a university name redacted for the blind review], as a part of the application 2031428-1. The detailed procedure to create these two message pools is explained in Supplementary Online Material (SOM).

**Human Evaluation**

Human evaluators assessed 1,200 long-form and 1,200 short-form messages selected from the message pools. The details of the selection procedure are provided in SOM. The selected messages were evaluated by a team of three human evaluators. Specifically, each message was independently assessed and classified by two evaluators, and in cases of disagreement, a third evaluator resolved the discrepancies. The inter-coder reliability among the evaluators was very high: Cohen's Kappa scores were .938 and .885 for long-form and short-form messages, respectively. The primary focus of this research lies in the capability of LLMs, which are designed to generate human-like assessments, in accurately replicating human evaluations of opinions on a contentious public health issue (Refer to SOM for further explanation).

**Machine Evaluation**

From the long-form messages assessed by human coders, we randomly selected 200 pro-vaccination, 200 anti-vaccination, and 100 neutral messages. Similarly, from the human-evaluated short-form messages, 200 pro-vaccination, 200 anti-vaccination, and 100 neutral messages were randomly selected. The current research refers to these refined groups of



messages as "machine evaluation sets." All messages were then evaluated by GPT 3.5. We used the model's latest version as of September 2023 (model name: gpt-3.5-turbo-0613). In order to compare the results from an extended number of iterations, we utilized an automated Python script based on OpenAI's commercial API (Application Programming Interface). The same tasks can be completed with ChatGPT by entering written prompts into its free web interface. This option is particularly advantageous for researchers seeking computational analysis of small or moderate-sized datasets who may lack technical knowledge, coding abilities, or financial resources, even though the API offers a more efficient, streamlined approach for evaluating a large amount of messages without the need for repetitive manual input.

For each message in a machine evaluation set, a prompt was created and presented to the language model. The prompt included instructions, the content of the message, and the coding scheme, as presented in SOM. The instructions commanded the model to classify a message into one of the five categories based on the coding scheme and explain its decision: ANTI (anti-vaccination), PRO (pro-vaccination), NEU (neutral), MIX (mixed), and IR (irrelevant). Except for minor formatting adjustments, the instructions and the coding scheme were identical to those provided to the human evaluators. Considering that identical prompts may yield varying responses due to the probabilistic nature of language models [25], we gathered 20 response instances for each message and thus a total of 20,000 response instances. It was done by initiating a new "chat" with the model, sending the prompt in the chat, receiving and storing its response, and repeating the process 20 times for each message.

The language model's decision for each message, termed "machine decision," was determined by randomly selecting $m$ out of the 20 response instances with replacement and identifying the majority of answers. This approach considers the 20 response instances as a



sample of possible evaluations generated by the model for a given message. To compare accuracy across different numbers of response instances, we varied $m$ among values of 1, 3, 5, 7, 9, and 11. For example, the case of $m = 3$ emulates a scenario in which a user generates three response instances and determines the majority among them. If a message received categorizations of ANTI, ANTI, and PRO with $m = 3$, the machine decision would be determined as ANTI. When there was a tie, one additional response instance was randomly selected until the tie was resolved. $m = 1$ corresponds to "one-shot" determination, where one instance was generated and considered as the machine decision.

For each message, we iterated the random selection and majority determination process 1,000 times. After each iteration, a value referred to as "human-machine concurrence" was recorded as 1 if the machine decision matched the human evaluation of the message; otherwise, it was recorded as 0. This variable was then averaged across all iterations, resulting in a value referred to as "accuracy." This accuracy value reflects the model's accuracy for a specific message. Furthermore, we computed the average accuracy across all the messages within a machine evaluation set, denoted as $K_m$. This provides an assessment of the model's overall accuracy for the messages within that particular set.

## Results

Average accuracy based on 20 response instances ($K_{20}$) was notably high for anti- and pro-vaccination messages. When considering anti- and pro-vaccination messages together, $K_{20}$ was .816 ($SE = .018$) for long-form and .748 ($SE = .020$) for short-form messages. These results are particularly noticeable considering that machine evaluation was conducted without any tailored pre-training or fine-tuning specific to HPV vaccination discussion. This highlights the large language model's capability and efficiency in distinguishing stances toward vaccination.



Specifically for anti-vaccination messages, average accuracy was even higher: $K_{20}$ achieved .882 ($SE$ = .021) and .773 ($SE$ = .027) for long-form and short-form messages, respectively.

Importantly, however, the language model exhibited lower accuracy for pro-vaccination messages than anti-vaccination ones in the long form: $K_{20}$ was .882 ($SE$ = .021) for anti-vaccination messages, whereas it was .750 ($SE$ = .029) for pro-vaccination ones. The difference was statistically significant (Mann-Whitney $U$ = 22779, $p$ = .005). While the level of statistical significance diminishes as $m$ decreases, a pattern linked to increasing variability induced by fewer response instances for majority determination (See SOM for the complete test results), the consistent gap in average accuracy can be observed in Figure 1. For short-form messages, however, the difference in average accuracy between anti- and pro-vaccination messages was not statistically significant even with 20 response instances ($U$ = 21038, $p$ = .324).

Furthermore, average accuracy for neutral messages was relatively low: $K_{20}$ was merely .540 ($SE$ = .045) for long-form and .541 ($SE$ = .042) for short-form messages. These outcomes were significantly lower than those of anti-vaccination messages (long-form: $U$ = 15455.5, $p$ < .001; short-form: $U$ = 14135, $p$ < .001) and pro-vaccination messages (long-form: $U$ = 13615, $p$ < .001; short-form: $U$ = 13615, $p$ < .001). As visualized in Figure 1, this decline in accuracy for neutral messages was consistent across different response instance counts and formats (See SOM for all test results).

It is worth noting that considerable levels of accuracy could be achieved with only a few response instances, underscoring the efficiency of using the language model for sentiment analysis, as shown in Table 1. Even when employing a relatively small number of instances, such as $m$ = 1 and 3, the average accuracy did not experience a severe decline. For instance, across anti-vaccination, pro-vaccination, and neutral content in long-form, average accuracy with three



instances ($K_3$) reached 95.2%, 97.3%, and 93.4% of those of 20 instances, respectively. The average accuracy of one-shot determination ($K_1$) also achieved 87.2%, 93.0%, and 86.1% of $K_{20}$ for anti, pro, and neutral content in long-form, respectively. A similar pattern was found for short-form messages (Table 1). Average accuracy increased with the number of response instances used for majority determination, albeit with diminishing returns as visualized in Figure 1. $K_{11}$ surpassed 98% of $K_{20}$ across all evaluation sets.

## Discussion

The present research underscores evidence of the potential of LLMs as tools for sentiment analysis of social media content about socially contentious public health issues. The findings demonstrate that ChatGPT powered by GPT 3.5 exhibits considerable accuracy in evaluating messages related to HPV vaccination. However, the research also highlights that the accuracy of LLMs can significantly fluctuate depending on the message content and format. The findings reveal that GPT 3.5 displays lower accuracy in identifying pro-vaccination messages compared with anti-vaccination ones for long-form messages. The language model also encountered difficulties in accurately replicating human evaluation decisions for neutral messages. Additionally, the model's accuracy was lower for short-form messages than long-form ones, differing from findings in a study on political texts[18].

These discrepancies pose substantial challenges in the practical application of the language model, necessitating additional techniques and procedures to assess, mitigate, or compensate for the inconsistencies. This may also involve new approaches to crafting instructions and coding schemes that enhance machine accuracy for pro-vaccination messages, neutral content, or shorter messages. Researchers must be aware of the characteristics and limitations inherent to LLMs to ensure the reliability and validity of research outcomes.



This research is not without limitations. Most of all, the present study primarily focused on examining the accuracy of a widely used language model, in evaluating vaccine-related messages from the two major social media platforms. Additional discussions on the limitations are provided in SOM.

## Disclosure of Funding and Conflict of Interest

The authors declare no conflicts of interest related to this study. This research received no specific grant from any funding agency in the public, commercial, or not-for-profit sectors.

## Data Availability

Supplementary information can be accessed via https://osf.io/ju39c



Table 1. Machine Accuracy of Sentiment Evaluation by the Number of Response Instances for

Majority Determination

| | Facebook (Long format) | | | | | | | | | |
|---|---|---|---|---|---|---|---|---|---|---|
| $m$ | ANTI ($n = 200$) | | PRO ($n = 200$) | | NEU ($n = 100$) | | ANTI & PRO ($n = 400$) | | All ($N = 500$) | |
| | $K_m$ (SE) | $K_m / K_{20}$ | $K_m$ (SE) | $K_m / K_{20}$ | $K_m$ (SE) | $K_m / K_{20}$ | $K_m$ (SE) | $K_m / K_{20}$ | $K_m$ (SE) | $K_m / K_{20}$ |
| 1 | .770 (.019) | 87.2% | .697 (.025) | 93.0% | .468 (.031) | 86.1% | .734 (.016) | 89.9% | .681 (.015) | 89.4% |
| 3 | .840 (.020) | 95.2% | .729 (.026) | 97.3% | .508 (.037) | 93.4% | .785 (.017) | 96.1% | .729 (.016) | 95.8% |
| 5 | .860 (.020) | 97.4% | .738 (.027) | 98.4% | .521 (.040) | 96.0% | .799 (.017) | 97.9% | .743 (.017) | 97.6% |
| 7 | .866 (.020) | 98.2% | .741 (.028) | 98.9% | .527 (.041) | 97.1% | .804 (.017) | 98.5% | .748 (.017) | 98.3% |
| 9 | .872 (.020) | 98.8% | .744 (.028) | 99.2% | .531 (.042) | 97.7% | .808 (.018) | 99.0% | .752 (.017) | 98.8% |
| 11 | .875 (.021) | 99.2% | .746 (.028) | 99.5% | .535 (.043) | 98.4% | .811 (.018) | 99.3% | .755 (.017) | 99.2% |
| 20 | .882 (.021) | - | .750 (.029) | - | .540 (.045) | - | .816 (.018) | - | .761 (.394) | - |
| | Twitter (Short format) | | | | | | | | | |
| $m$ | ANTI ($n = 200$) | | PRO ($n = 200$) | | NEU ($n = 100$) | | ANTI & PRO ($n = 400$) | | All ($N = 500$) | |
| | $K_m$ (SE) | $K_m / K_{20}$ | $K_m$ (SE) | $K_m / K_{20}$ | $K_m$ (SE) | $K_m / K_{20}$ | $K_m$ (SE) | $K_m / K_{20}$ | $K_m$ (SE) | $K_m / K_{20}$ |
| 1 | .679 (.023) | 87.9% | .675 (.024) | 93.3% | .448 (.026) | 82.9% | .677 (.017) | 90.5% | .631 (.015) | 89.3% |
| 3 | .735 (.025) | 95.1% | .702 (.027) | 97.0% | .498 (.033) | 92.0% | .718 (.018) | 96.1% | .674 (.017) | 95.4% |
| 5 | .751 (.026) | 97.2% | .711 (.028) | 98.3% | .514 (.036) | 95.1% | .731 (.019) | 97.7% | .687 (.017) | 97.3% |
| 7 | .757 (.026) | 98.0% | .713 (.028) | 98.6% | .520 (.037) | 96.1% | .735 (.019) | 98.3% | .692 (.018) | 98.0% |
| 9 | .762 (.027) | 98.6% | .717 (.028) | 99.1% | .525 (.039) | 97.0% | .740 (.019) | 98.9% | .697 (.018) | 98.6% |
| 11 | .765 (.027) | 99.1% | .719 (.029) | 99.4% | .532 (.040) | 98.4% | .742 (.020) | 99.2% | .700 (.018) | 99.1% |
| 20 | .773 (.027) | - | .723 (.029) | - | .541 (.042) | - | .748 (.020) | - | .707 (.018) | - |

*Note. $m$* is the number of response instances generated. When $m > 1$, a machine decision was determined by the majority rule among $m$ response instances. $m = 1$ corresponds to one-shot evaluations without majority determination. $K_m$ is machine accuracy averaged across $n$ messages when $m$ response instances were generated to determine machine decision. ANTI, PRO, and NEU indicate human-evaluated anti-vaccination, pro-vaccination, and neutral messages.



Figure 1. Machine Accuracy of Sentiment Evaluation by the Number of Response Instances for

Majority Determination

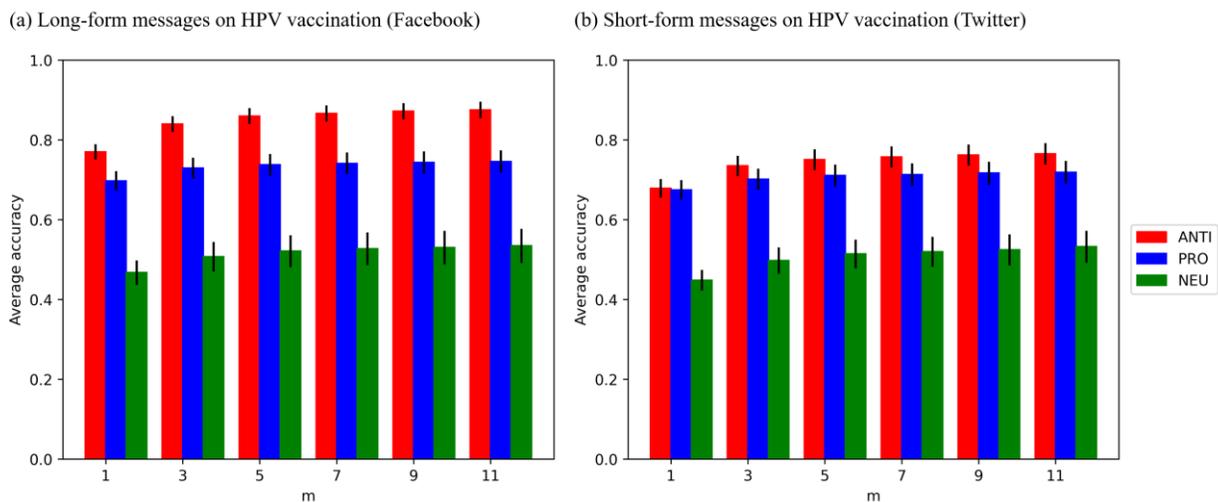

(a) Long-form messages on HPV vaccination (Facebook)    (b) Short-form messages on HPV vaccination (Twitter)

*Note*. ANTI, PRO, and NEU indicate human-evaluated anti-vaccination, pro-vaccination, and neutral messages. $m$ is the number of response instances generated. When $m > 1$, a machine decision was determined by the majority rule among m response instances. $m = 1$ corresponds to one-shot evaluations without majority determination. Bars indicate average accuracy, and error bars indicate mean ± s.e.m.